\documentclass{PoS}

\PoS{PoS(LAT2005)350}
\newcommand{\pslash}{p\kern-1ex /}
\newcommand{\les}{\stackrel{<}{{}_{\sim}}}

\newcommand{\VEV}[3]{\left\langle #1\left| #2 \right| #3\right\rangle}

\newcommand{\nn}{\nonumber}
\newcommand{\ovl}[1]{\overline{#1}}

\newcommand{\Tr}{{\rm Tr}}

\title{$K\to\pi\pi$ from electroweak penguins in $N_f=2$ domain-wall QCD}

\ShortTitle{$K\to\pi\pi$ from electroweak penguins in $N_f=2$ 
domain-wall QCD }

\author{\speaker{Jun Noaki}\thanks{I thank the RBC collaboration for
        allowing me to use the data generated for their projects.
        This work is done under the support by JSPS Postdoctoral Fellowships 
	for Research Abroad.}\\
	School of Physics and Astronomy\\
        University of Southampton\\
	Southampton SO17 1BJ, England.\\
        E-mail: \email{noaki@phys.soton.ac.uk}}

\abstract{We present the calculation of $K\to\pi\pi$ matrix elements of 
electroweak penguin operators {\it i.e.} $Q_7$ and $Q_8$.
In the numerical simulation, we use the gauge configurations generated 
by the combination of $N_f=2$ domain-wall fermions and DBW2 gauge action.
From $K\to\pi$ and $K\to 0$ matrix elements on the lattice, we construct
$K\to\pi\pi$ matrix elements at next-to-leading order in the chiral
expansion by using recent analytic results.
Renormalization factor of these matrix elements are obtained by the 
non-perturbative renormalization technique in RI/MOM scheme. 
Brief discussion based on our results are made as well as 
a comparison with previous works.}

\FullConference{XXIIIrd International Symposium on Lattice Field Theory\\
		 25-30 July 2005\\
		 Trinity College, Dublin, Ireland}
\begin{document}

\section{Introduction}

In the efforts toward the theoretical treatment of the non-leptonic 
kaon decay, $\epsilon'/\epsilon$, the direct CP violation parameter of 
$K\to\pi\pi$, is one of the important focuses.
This quantity is approximated to a linear combination of the $K\to\pi\pi$ 
matrix elements of the local operators $Q_6$ and $Q_8$ in the 
$\Delta S=1$ effective hamiltonian  
$H_W^{(\Delta S=1)}= \sum_{i=1}^{10}W_i(\mu)Q_i(\mu)$
(where $W_i(\mu)$ is the Wilson coefficients obtained by perturbative 
calculations). 
However, calculations in lattice QCD (e.g.~\cite{CPPACS2003,RBC2003})
resulted in unacceptable central values of $\epsilon'/\epsilon$ due to 
a small magnitude of $\VEV{\pi\pi_{\ (I=0)}}{Q_6}{K}$ compared to 
$\VEV{\pi\pi_{\ (I=2)}}{Q_8}{K}$.
In these works, $K\to\pi\pi$ matrix elements are obtained through
the low energy constants (LECs) extracted from $K\to\pi$ and $K\to 0$ 
matrix elements on the lattice. 
Because this method is based on the lowest order (LO) chiral perturbation
theory (ChPT), results could change significantly by taking the 
next-to-leading order (NLO) effects into account. 

For electroweak penguin operators $Q_{\rm ewp}=Q_{7,8}$, there are 
limited number of NLO ChPT operators due to their lower mass dimension
than operators in other classes. By recent efforts
~\cite{CiriglianoGolowich2000,GoltermanPallante2001,LaihoSoni2005},
there arose a prospection that the LO method can be extended 
to NLO for $Q_{\rm ewp}$.
In particular, the authors of Ref.~\cite{LaihoSoni2005} have shown that 
it is possible to extract all LECs needed to construct $K\to\pi\pi$ 
at NLO from $K\to\pi$ and $K\to 0$ matrix elements in the framework 
of partially quenched ChPT (PqChPT).
Following the line, we calculate $K\to\pi\pi$ matrix elements of 
electroweak penguins on dynamical gauge configurations.

\section{Numerical simulation and analysis}

We use $N_f=2$ gauge configurations generated by the RBC 
Collaboration~\cite{RBC_DYN2004} using domain-wall fermions with 
$L_s=12$ and $M_5=1.8$ and DBW2 gauge action with $\beta=0.80$ 
in the $16^3\times 32$ box. 
There are ensembles with $m_{\rm sea} =0.02$, $0.03$ and $0.04$,
each of which 94 configurations are available.
$a^{-1}=1.69(5)$ GeV and $am_{\rm res} = 0.00137(5)$ have been
obtained in Ref.~\cite{RBC_DYN2004}.

Electroweak penguin operators are defined as 
\begin{eqnarray}
Q_7= (\bar{s}_aL_\mu d_a)\sum_{q = u,d,s}e_q(\bar{q}_bR_\mu q_b), \ \ \ 
Q_8= (\bar{s}_aL_\mu d_b)\sum_{q = u,d,s}e_q(\bar{q}_bR_\mu q_a),
\end{eqnarray}
where $L_\mu=(1-\gamma_5)\gamma_\mu$ and $R_\mu=(1+\gamma_5)\gamma_\mu$,
and $(e_u, e_d, e_s)=(2/3,-1/3,-1/3)$. Roman indices represent color.
Each operator is divided into two contributions with $\Delta I=1/2$ and 
$3/2$ as $Q_{\rm ewp}=Q_{\rm ewp}^{(1/2)} + Q_{\rm ewp}^{(3/2)}$.
We calculate correlation functions with the wall pseudo-scalar 
operator $P(t)$ and the point operator $Q_{\rm ewp}({\bf x}, t)$ for
each $m_{\rm sea}$. With valence quark masses 
$m_{\rm val}= 0.01,0.02,0.03,0.04,0.05$, matrix elements are 
computed as a plateau in the $t$-dependence of the ratios
\begin{eqnarray}
\VEV{\pi^+}{Q_{\rm ewp}^{(\rm latt)}}{K^+} &=& 
\frac{\sum_{\bf x}\VEV{0}{P(t_f)Q_{\rm ewp}({\bf x}, t)P^\dagger(t_i)}{0}}
     {\VEV{0}{P(t_f)P^\dagger(t_i)}{0}}\times 2m_{PS},\\
\VEV{0}{Q_{\rm ewp}^{(\rm latt)}}{K^0} &=&
\frac{\sum_{\bf x}\VEV{0}{Q_{\rm ewp}({\bf x}, t)P^\dagger(t_i)}{0}}
     {\VEV{0}{P(t)P^\dagger(t_i)}{0}}\times 
     \sqrt{\frac{2m_{PS'}}{V}\cdot{\rm Amp}^{(PS')}},
\end{eqnarray}
where $m_{PS}$ and $m_{PS'}$ are pseudo-scalar meson masses for 
degenerate and non-degenerate quark, respectively.
${\rm Amp}^{(PS')}$ is the amplitude of the $PP$-correlator.
Locations of sources are $(t_i,t_f)=(5,27)$ and we use 
$t=14$--$17$ for the fit range.

One may write $Q_{\rm ewp}$ as an expansion of ChPT operators
with $(8_L, 8_R)$ representation:
\begin{eqnarray}
Q_{\rm ewp} = \alpha_{88}\left(\Sigma_{13}\Sigma_{21}^\dagger\right)
+\sum_{i=1}^{6}c_i{\cal O}_i^{\rm (ewp)}+O(p^4),
\end{eqnarray}
where ${\cal O}_i^{\rm (ewp)}$ are the NLO operators.
Functions to which we fit our numerical results are~\cite{LaihoSoni2005}
\begin{eqnarray}
\VEV{\pi^+}{Q_{\rm ewp}^{(1/2)}}{K^+} &=& \alpha_{88}\Biggl[
\frac{1}{2\pi^2f^4}
\left\{
 m_{PS}^2\ln\frac{m_{PS}^2}{\mu^2}
-4(m_{PS}^2+m_{SS}^2)\ln\left( \frac{m_{PS}^2+m_{SS}^2}{2\mu^2}\right)
+m_{PS}^2
\right\}\nn\\
& &
-\frac{16}{f^3}(f_{PS}-f) + \frac{8}{f^2}\Biggr]
+\xi^{(1/2)}\frac{4}{f^2}m_{PS}^2 
+ c^r_6 \frac{32}{f^2}m_{SS}^2 +O(m_{PS}^4),\label{kpi1hf}\\
\VEV{\pi^+}{Q_{\rm ewp}^{(3/2)}}{K^+} &=& \alpha_{88}\Biggl[
\frac{-1}{2\pi^2f^4}
\left\{
 m_{PS}^2\ln\frac{m_{PS}^2}{\mu^2}
+8(m_{PS}^2+m_{SS}^2)\ln\left( \frac{m_{PS}^2+m_{SS}^2}{2\mu^2}\right)
+m_{PS}^2
\right\}\nn\\
& &
-\frac{8}{f^3}(f_{PS}-f) + \frac{4}{f^2}\Biggr]
+\xi^{(3/2)}\frac{4}{f^2}m_{PS}^2 
+ c^r_6 \frac{16}{f^2}m_{SS}^2 +O(m_{PS}^4),\label{kpi3hf}\\
\VEV{0}{Q_{\rm ewp}^{(1/2)}}{K^0}&=& \frac{\alpha_{88}}{4\pi^2f^3}\Biggl[
2m_{PS'}^2\ln\frac{m_{PS'}^2}{\mu^2}-m_{PS}^2\ln\frac{m_{PS}^2}{\mu^2}
-(2m_{PS'}^2-m_{PS}^2)\ln\left(\frac{2m_{PS'}^2-m_{PS}^2}{\mu^2}\right)\nn\\
& &+2m_{sS}^2\ln\frac{m_{sS}^2}{\mu^2}+2m_{uS}^2\ln\frac{m_{uS}^2}{\mu^2}
\Biggr]
-\frac{8c_4^r}{f}(m_{PS'}^2-m_{PS}^2)+ O(m_{PS}^4)\label{kto0},
\end{eqnarray}
where $m_{SS}^2=m_{PS}^2{}_{(m_{\rm val}=m_{\rm sea})}$.
For a set of non-degenerate valence quark masses such that
$m_{\rm val1}< m_{\rm val2}$, we define 
$m_{sS}^2=m_{PS'}^2{}_{(m_{\rm val1}= m_{\rm sea})}$ and 
$m_{uS}^2=m_{PS'}^2{}_{(m_{\rm val2}= m_{\rm sea})}$.
Pseudo-scalar decay constant $f_{PS}$ computed with degenerate 
quark masses is extrapolated to $f$ in the chiral limit.
After cancelling the 1-loop divergences from the first term,
there remain constant coefficients $c_i^r$ as LECs for NLO.
Note that, in above functions, the number of LECs reduced from seven 
to five: 
$\left\{\alpha_{88},\, \xi^{(1/2)},\, \xi^{(3/2)},\, c_4^r,\, c_6^r\right\}$,
where $\xi^{(1/2)}= -c_1^r + c_2^r + 2c_3^r +10c_4^r +8c_5^r$ and 
$\xi^{(3/2)}= -c_1^r - c_2^r + 4c_4^r +4c_5^r$.
One can easily check that these LECs successfully construct $K\to\pi\pi$ 
matrix elements at NLO~\cite{LaihoSoni2005}.

\section{Non-perturbative renormalization}

Before computing $K\to\pi\pi$ matrix elements, we renormalize 
lattice values non-perturbatively in RI/MOM scheme~\cite{RIMOM1995}.
Numerical simulation for this step is carried out on 48,102 and 67 
configurations for $m_{\rm sea}=0.02,0.03$ and $0.04$, respectively.
The main procedure is based on the method in Ref.~\cite{RBC2003}.
We consider the renormalization factor of electroweak penguin $Z$ as a 
$2\times 2$ matrix to solve the mixing between $Q_7$ and $Q_8$. 
The renormalization condition is
\begin{eqnarray}
Z_q^{-2}Z_{ij}\Gamma^{\rm latt}_{Q_j}(p_1,p_2)= \Gamma^{\rm tree}_{Q_i},
\label{renorm_condition}
\end{eqnarray}
where $\Gamma^{\rm latt}_{Q_i}$ is the amputated Green's function 
on the lattice in the momentum space. 

In the case of $\Delta I=1/2$, there is a contribution from the 
disconnected diagram in which an one loop contraction of $Q_i$ exchanges 
gluons with the spectator. 
For its reasonable implementation, we assign two momenta $p_1$ and $p_2$ 
which satisfy $|p_1|=|p_2|=|p_1-p_2|$ to each part of diagram so that
all relevant momenta have same magnitude. In terms of lattice momentum 
$p_{\rm latt}\cdot 16/(2\pi)= (n_x, n_y, n_z, n_t)$,
four sets $(1,1,1,1),(0,2,2,0),(1,1,2,2),(2,2,2,2)$ and 4,8,8,4 
equivalent partners to each are employed. 
We average results over the sets of momenta with same magnitude. 

Another issue for $\Delta I=1/2$ is the mixing with lower dimension 
operators. As discussed in Ref.~\cite{RBC2003}, this mixing with 
the two major operators can be subtracted as 
\begin{eqnarray}
Q^{(1/2) {\rm subt}}_i = Q^{(1/2)}_i + c^1_i\cdot (\bar{s}d)
+c^2_i\cdot\bar{s}(\overleftarrow{D}_s + \overrightarrow{D}_d)d
\label{subtraction}
\end{eqnarray}
by determining the mixing coefficients $c_{7,8}^1$ and $c_{7,8}^2$.
As a result of the procedure described in in Ref.~\cite{RBC2003},
we find the effect of subtraction is $\les 1\%$ for the diagonal 
elements and $\les 6\%$ for the off-diagonal ones.

\begin{figure}
\begin{center}
\includegraphics[width=6.4cm,clip]{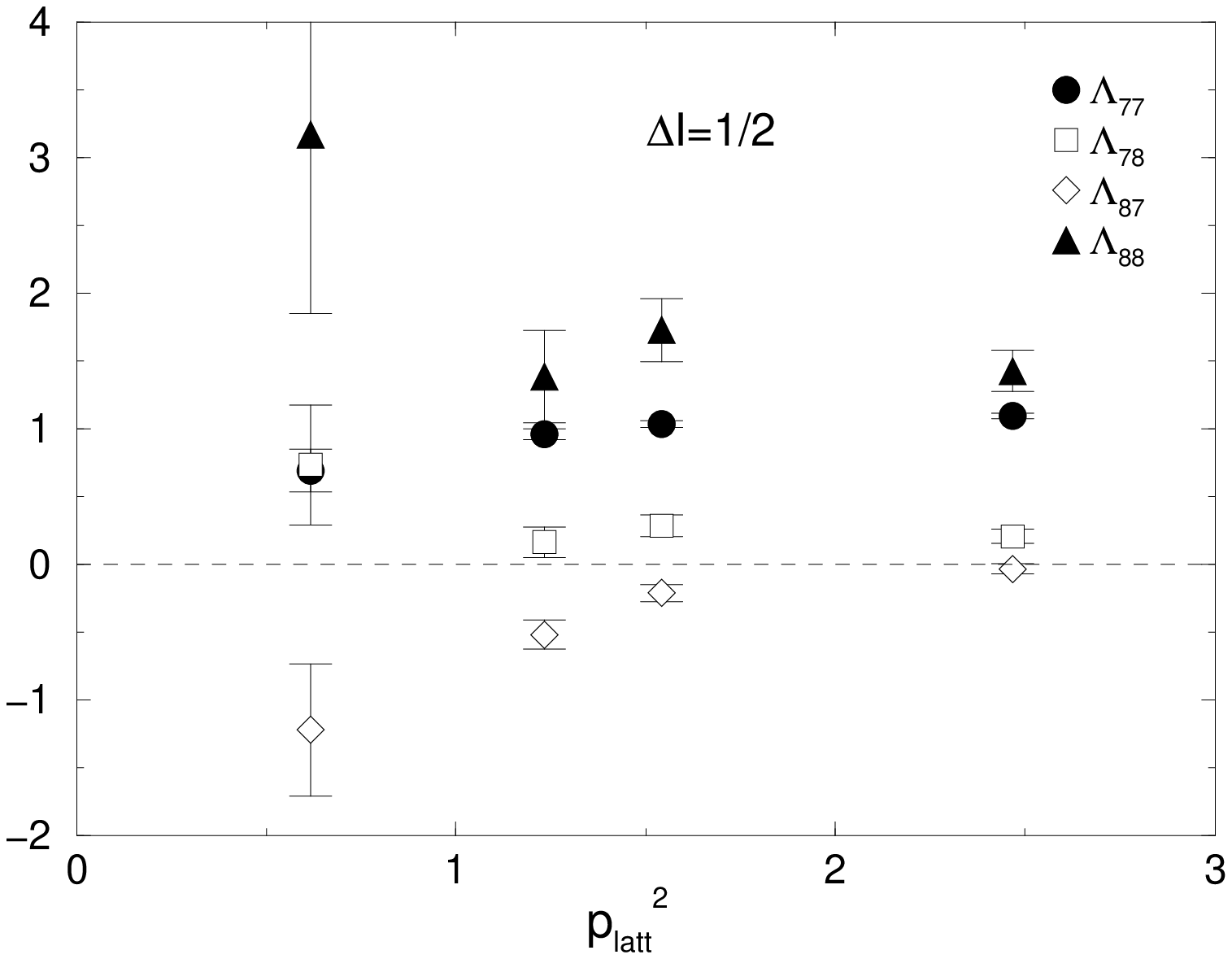}
\hspace{0.5cm}
\includegraphics[width=6.4cm,clip]{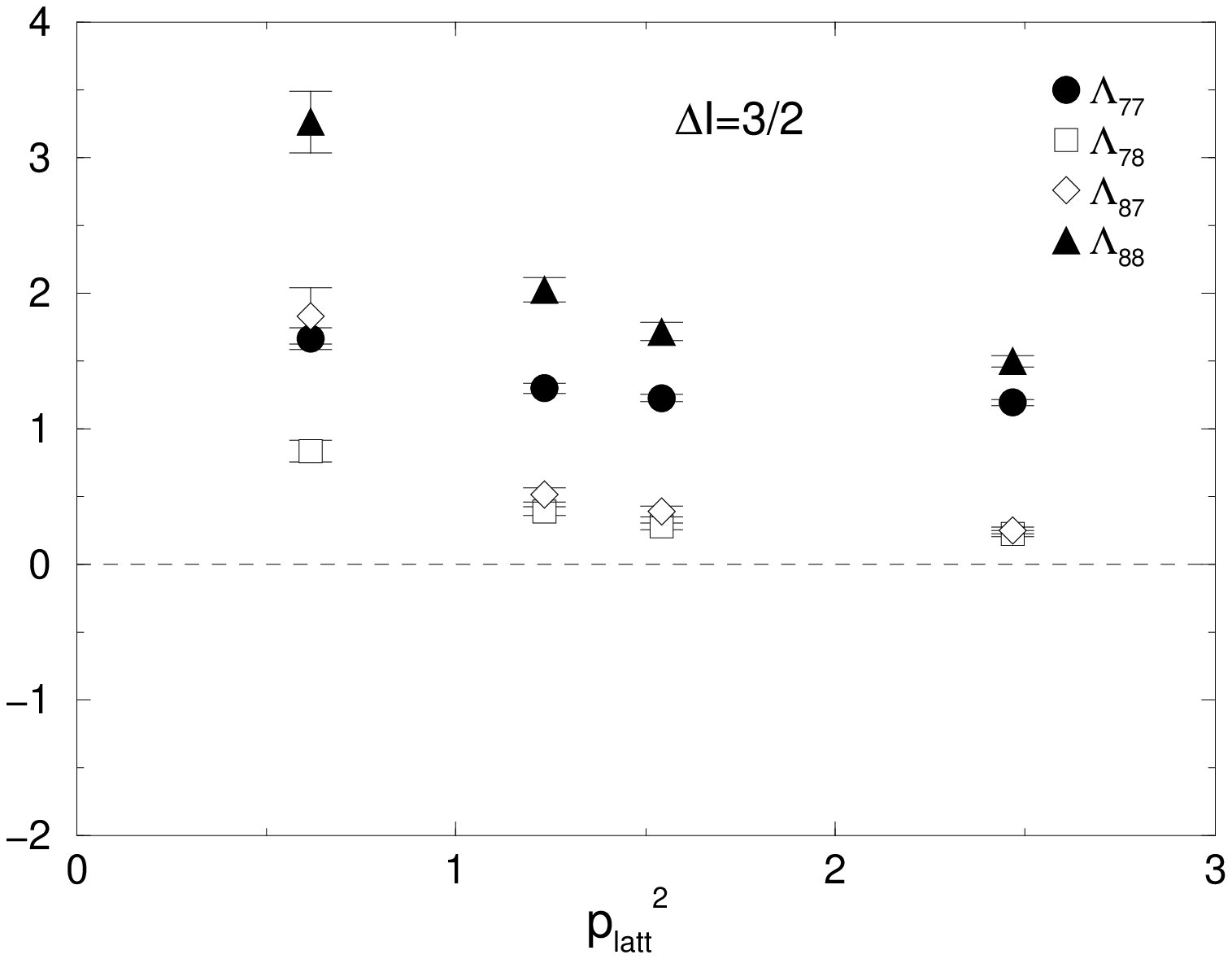}
\caption{$\Lambda_{ij}$ for $\Delta I=1/2$ (left) and $3/2$
 (right) in the chiral limit as a function of $p_{\rm latt}^2$. 
 In each panel, diagonal (off-diagonal) elements are indicated by filled 
 (open) symbols.}
\label{NPRresult}
\end{center}
\end{figure}

From (\ref{renorm_condition}), the renormalization factor 
$Z_q^{-2}Z_{ij}$ is the inverse of 
$\Lambda_{ij}\equiv \Tr[\Gamma^{\rm latt}_{Q_i}\Gamma^{\rm tree}_{Q_j}]$. 
After the chiral extrapolation using the data with 
$m_{\rm sea}=m_{\rm val}$, we show $\Lambda_{ij}$ in 
Figure~\ref{NPRresult} as a function of $p_{\rm latt}^2$ for 
$\Delta I=1/2$ (left panel) and $3/2$ (right panel). 
Momentum dependences for $p_{\rm latt}^2\les 1.5$ imply the contamination 
of low energy effects. To avoid this, we employ the data at the 
largest momentum $(ap_{\rm latt})^2= 2.467$ for the rest of analysis. 
Given differences between two largest momentum are $\les 2 \sigma$ 
except for the $87$-elements for both $\Delta I$, we expect this 
contamination is under control. This small difference 
also moderates the worry that the $O(a^2)$ error is too large for 
the NPR technique to work correctly.

After obtaining $Z_{ij}$ by an alternative calculation of $Z_q$,
we move to $\ovl{\rm MS}$ with NDR through the perturbative 
procedure~\cite{RI2MS}. At $\mu=a^{-1}$, our preliminary results are
\begin{eqnarray}
Z_{\rm \ovl{MS},\ NDR}^{(\Delta I=1/2)}= 
\left[
\begin{array}{cc}
0.599(21) & -0.059(38)\\
0.010(30) &\ \ 0.557(93)\\
\end{array}
\right],\ \ \ 
Z_{\rm \ovl{MS},\ NDR}^{(\Delta I=3/2)}= 
\left[
\begin{array}{cc}
\ \ 0.564(22) & -0.060(15)\\
-0.121(21)&\ \ 0.550(30)\\
\end{array}
\right],
\end{eqnarray}
where only the statistical errors are considered.
While the diagonal elements are $\approx 40\%$ smaller than the 
perturbative results~\cite{Aokietal1999}, similar results have been 
obtained by RI/MOM in Ref.~\cite{Becirevic_etal2004}.

\section{Results and discussion}

\begin{figure}
\begin{center}
\includegraphics[width=6.9cm,clip]{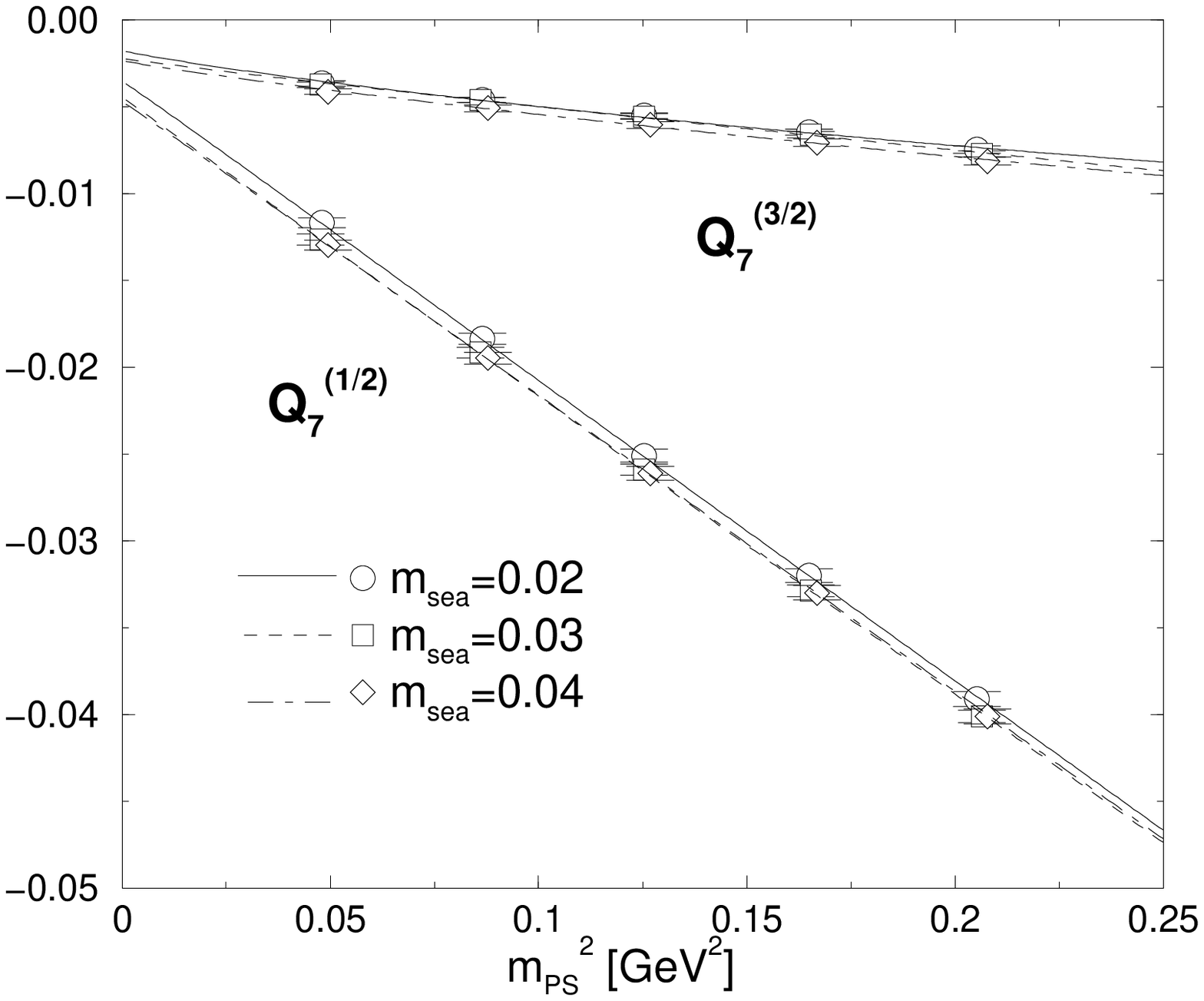}
\includegraphics[width=6.9cm,clip]{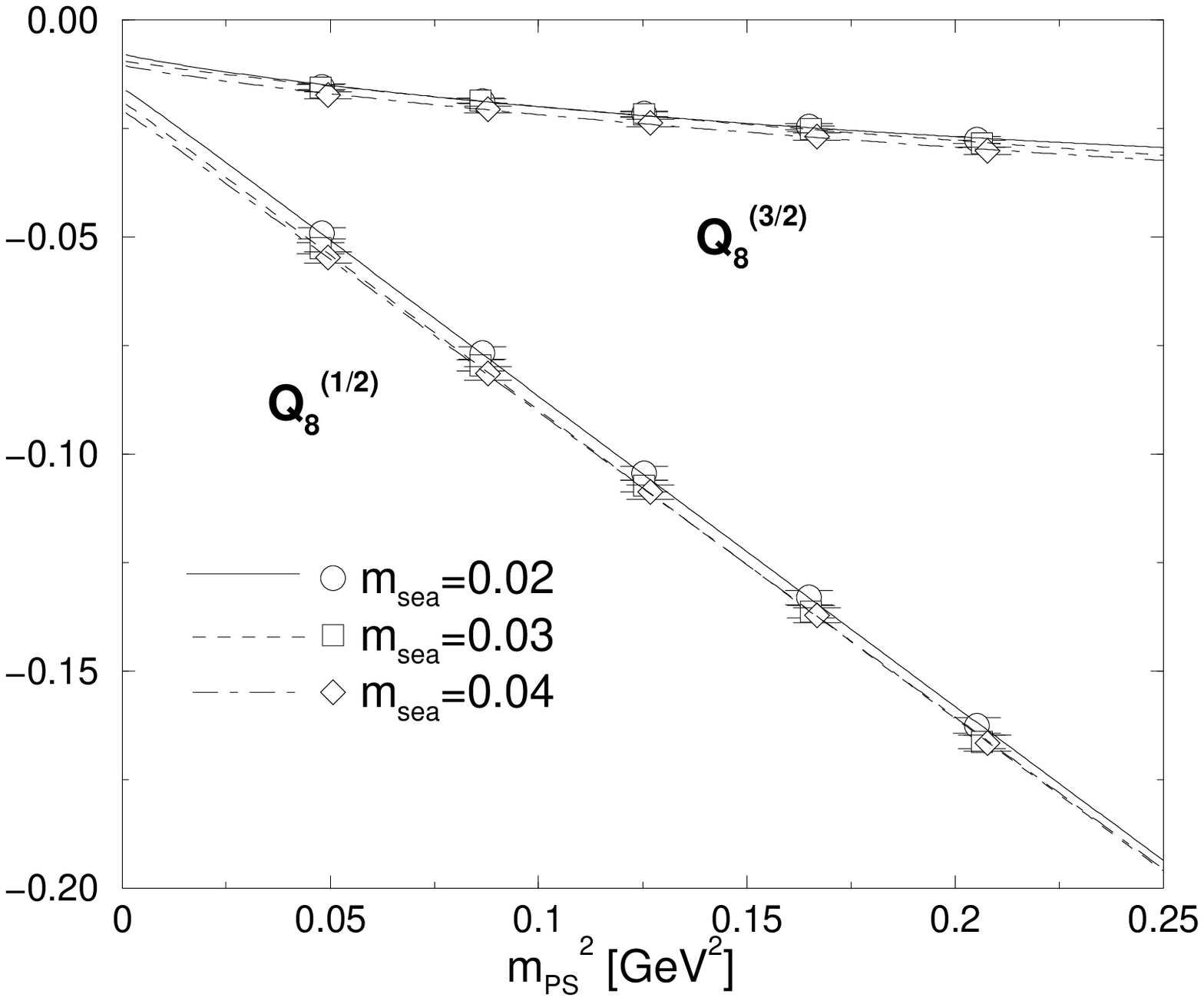}
\caption{Renormalized $K\to\pi$ matrix elements of $Q_7$
 (left) and $Q_8$ (right) as a function of $m_{PS}^2$ [${\rm GeV}^2$]. 
 In each panel, matrix elements with $\Delta I=1/2$ and $3/2$ are 
presented for $m_{\rm sea}=0.02$ (circles), $0.03$ (squares) and $0.04$
(diamonds). Fit curves are obtained from simultaneous fits to the 
PqChPT results.} 
\label{Fit_fig}
\end{center}
\end{figure}

We fit the renormalized matrix elements to (\ref{kpi1hf}), (\ref{kpi3hf}) 
and (\ref{kto0}) simultaneously to ensure common values of $\alpha_{88}$ 
and $c_6^r$. This fit is repeated independently for each $m_{\rm sea}$.
Using all data points (5 points for $K\to\pi$ and 10 points for $K\to 0$),
we find reasonable quality of fit with $0.2< \chi^2/{\rm dof} < 0.7$.
In Figure~\ref{Fit_fig}, we plot $K\to\pi$ matrix elements of $Q_7$
(left panel) and $Q_8$ (right panel) and their fit curves for each 
$m_{\rm sea}$.  

\begin{figure}
\begin{center}
\includegraphics[width=6.6cm,clip]{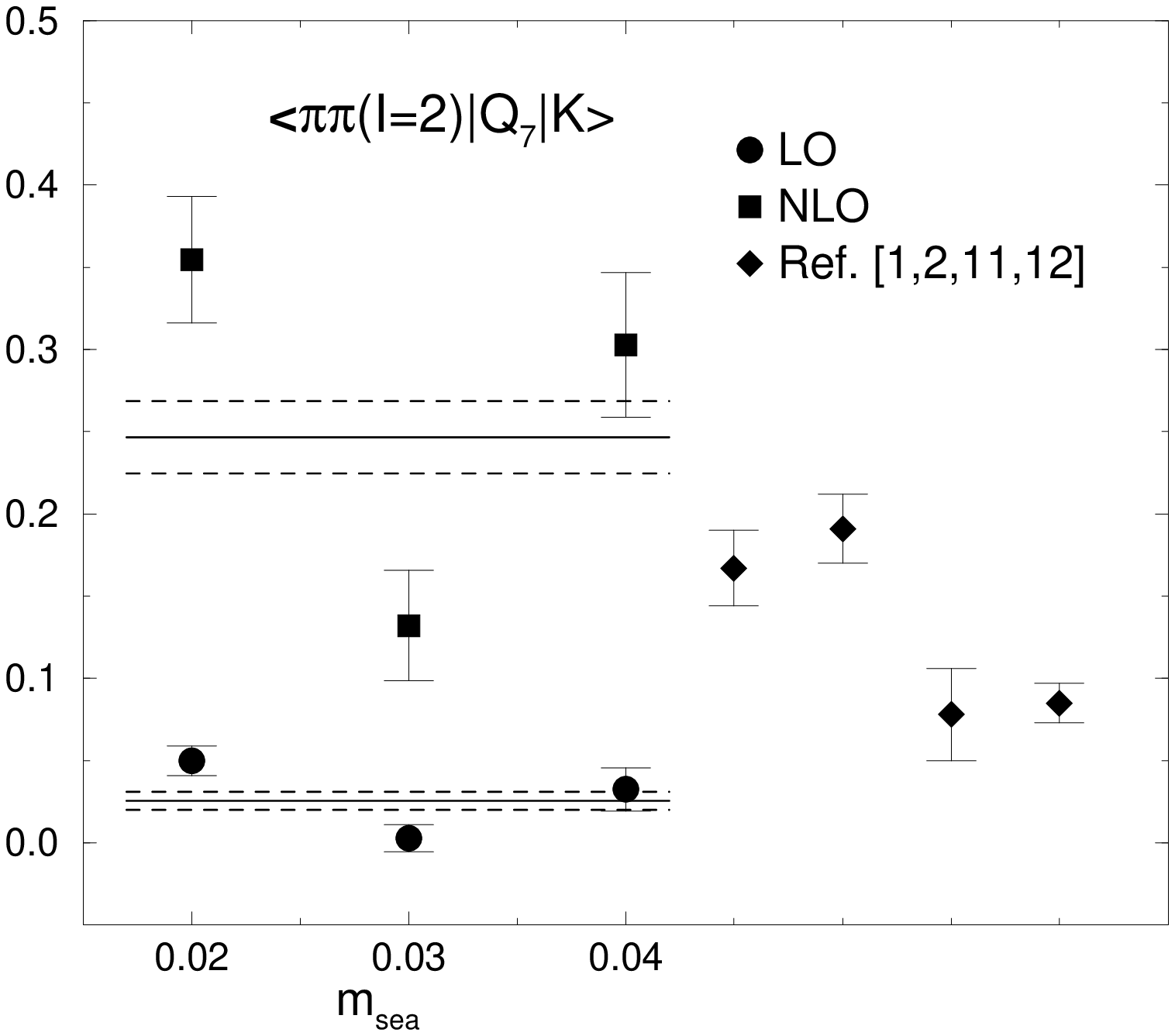}
\hspace{0.5cm}
\includegraphics[width=6.6cm,clip]{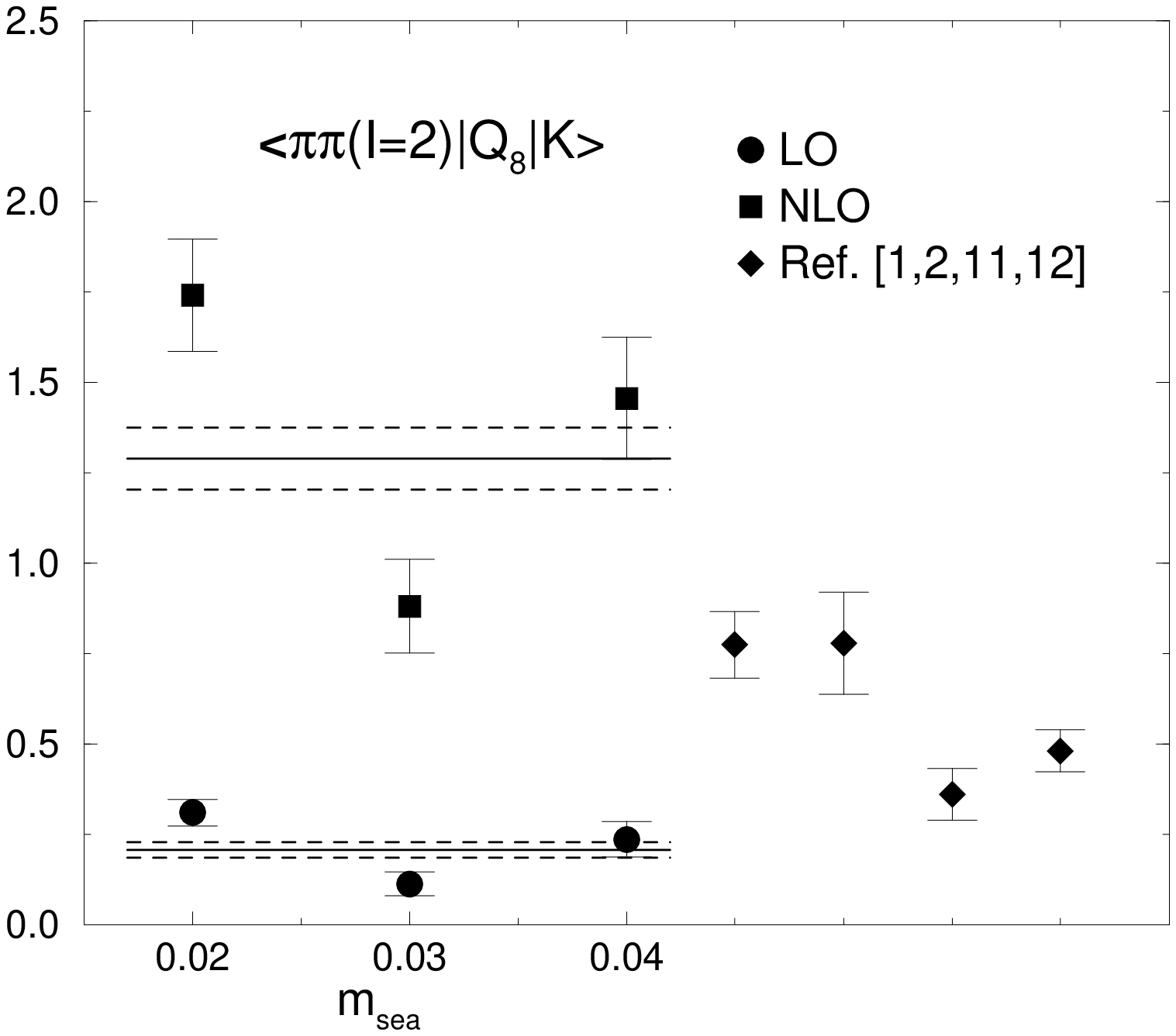}
\caption{Results of $K\to\pi\pi$ matrix elements with $\Delta I=3/2$ 
for $Q_7$ (left) and $Q_8$ (right) in the physical unit $\rm GeV^3$.
Averages over $m_{\rm sea}$'s in LO (circles) and NLO (squares) 
are indicated by horizontal lines. Filled diamonds are 
the results from previous works~\cite{CPPACS2003},~\cite{RBC2003},
~\cite{Donini_etal1999} and~\cite{Boucaud_etal2004}, from left to right.}
\label{KPPresult}
\end{center}
\end{figure}

By using LECs from each $m_{\rm sea}$ and physical meson masses 
and decay constants, we compute $K\to\pi\pi$ matrix elements at NLO. 
For comparison, we also obtain the LO value by neglecting higher 
order terms with same $\alpha_{88}$.
Figure~\ref{KPPresult} summarises the results of $K\to\pi\pi$ matrix 
elements in the physical unit with $\Delta I=3/2$ for $Q_7$ in left 
panel and $Q_8$ in right panel. 
The horizontal lines show the average over $m_{\rm sea}$'s with the 
weights of the errors. 
Results from previous quenched calculations
~Refs.\cite{CPPACS2003,RBC2003,Donini_etal1999,Boucaud_etal2004} 
are plotted in filled diamonds for comparison.  
We observe large enhancements from LO to NLO 
sandwiching all of quoted results both for $Q_7$ and $Q_8$.

Because NLO analysis of $\VEV{\pi\pi_{(I=0)}}{Q_6}{K}$ is not available
currently, it is not realistic to estimate $\epsilon'/\epsilon$ using
our NLO result of $\VEV{\pi\pi_{(I=2)}}{Q_8}{K}$.
Instead of that, we could use the LO result from the NLO analysis
in the estimation in the framework of LO.
It should be noted that we obtain consistent values to our NLO 
results by the LO method employed in Refs.~\cite{CPPACS2003,RBC2003}.
Therefore, assuming similar gap we have seen between NLO and LO in the 
quenched calculations, there should have been a large enhancements from 
higher order effects in Refs.~\cite{CPPACS2003,RBC2003}.
Then it is conceivable to obtain a fair value of $\epsilon'/\epsilon$ 
by using $\VEV{\pi\pi_{(I=2)}}{Q_8}{K}$ after removing large higher
order effects by the NLO analysis, which is only possible for unquenched 
calculation. However, more careful study of the systematics must be 
done for actual conclusion.

\newcommand{\NP}{Nucl.~Phys. }
\newcommand{\NPSup}{Nucl.~Phys.~{\bf B} (Proc.~Suppl.)}
\newcommand{\PL}{Phys.~Lett. }
\newcommand{\PRD}{Phys.~Rev. D }
\newcommand{\PRL}{Phys.~Rev.~Lett. }
\newcommand{\JHEP}{J.~High Energy~Phys. }

\end{document}